\documentclass[aps,twocolumn,showpacs]{revtex4}
\usepackage{graphicx}

\def\mg22{$^{22}Mg$}
\def\na22{$^{22}Na$}
\def\ne22{$^{22}Ne$}
\def\f19{$^{19}F$}
\def\xne19{$^{19}Ne$}
\def\yn13{$^{13}N$}
\def\c13{$^{13}C$}
\def\o18{$^{18}O$}
\def\zne18{$^{18}Ne$}

\def\gg{$\Gamma_\gamma$}

\begin{document}

\title{Comment on B.S. Davids {\em et al.}; \\
Proton-Decaying States in \mg22 and the Nucleosynthesis of \na22 in Novae}
\thanks{Work Supported by USDOE Grant No. DE-FG02-94ER40870.}

\author{Moshe Gai}
\altaffiliation{Permanent Address: Laboratory for Nuclear Science at Avery Point, 
University of Connecticut, 1084 Shennecossett Road, Groton, CT 06340-6097.}
\affiliation{Department of Physics, Room WNSL102, Yale University, \\
PO Box 208124, 272 Whitney Avenue, New Haven, CT 06520-8124. \\
\    \\
e-mail: moshe.gai@yale.edu, URL: http://www.phys.uconn.edu}

\begin{abstract}

B.S. Davids {\em et al.} make the explicit assumption that radiative widths 
of analog states in \ne22 and \mg22 are equal.  We demonstrate that this 
misapplication of iso-spin symmetry leads to very wrong results.  Considerations 
of elementary nuclear structure suggests that such an assumption can 
be inaccurate by a large factor (in \mg22), as is evident from a comparison 
with recent measurements of radiative width in \mg22. Estimates of radiative 
widths from analog transitions are common but often wrong (e.g. in \mg22) and 
should not be considered a useful tool in nuclear astrophysics.

\end{abstract}


\preprint{UConn-40870-00XX}

\maketitle

The (ab)use of isospin symmetry in Nuclear Astrophysics to estimate radiative widths 
is common and in this comment we consider one such (extreme) case.
B.S. Davids {\em et al.} \cite{Dav1} state that for the purpose of estimating the 
resonance strengths in \mg22 "we have assumed that they have the same radiative widths 
as their analog states in \ne22". Unfortunately similar statements already appeared 
in the literature and for example, previously B.S. Davids {\em et al.} stated 
\cite{Dav2}: "In some cases, widths of the analog state from the mirror nucleus 
\f19 have been measured, and we adopt these under the assumption that 
\gg(\xne19) = \gg(\f19)". And once again B.S. Davids {\em et al.} stated \cite{Dav3}: 
"Measurements of \gg  for analog states in the mirror nucleus \f19 can be found for four 
of the states and these have been adopted under the assumption that \gg(\xne19) = \gg(\f19)".
Since radiative widths are essential for nucleosynthesis in 
astrophysical environments, we chose to examine this assumption in detail. We demonstrate 
that in most cases the assumption of the equality of radiative widths of analog states has no 
physical justification, and most certainly in the case discussed in \mg22.

The total radiative width (\gg) depends on the reduced widths [B(E$\lambda$) or B(M$\lambda$)] 
as well as the phase space available for electromagnetic decays; i.e. the various states and 
the corresponding transitions available for decay with specific energies. 
Iso-spin symmetry does not imply that the total radiative width (\gg) of analog 
 states are equal, as assumed by B.S. Davids {\em et al.} \cite{Dav1,Dav2,Dav3}. Rather 
 it only makes predictions of the reduced widths. 
 Thus one must compare only the reduced widths for analog transitions and not the total 
 radiative widths of analog states. 
 
 This point is most 
 clear for analog states in self conjugate nuclei, where additional phase space is 
 available via allowed $\Delta$T = 1 electromagnetic transitions. Indeed in many cases and 
 most certainly in \mg22 we encounter electromagnetic decay with substantially different 
 energies and thus very different phase space that leads to very different radiative widths 
 of the analog states even if the reduced widths are equal. 

Moreover, even in the limit where iso-spin is an exact symmetry, in general the reduced widths [
B(E2) or B(M1)] of only $\Delta$T = $\pm$1 analog transitions are 
predicted to be equal (Rule 2 of Ref. \cite{War}). We emphasize 
that only for E1 transitions the analog transitions are predicted to 
be equal also for $\Delta$T = 0 transitions (Rule 3 of Ref. \cite{War}). And only strong or above 
average strength magnetic dipole (M1) transitions are expected to be approximately equal  
also for $\Delta$T = 0 (Rule 5 of Ref. \cite{War}). All too often and most 
certainly in \mg22 we deal with $\Delta$T = 0 E2 transitions or weak $\Delta$T = 0 
M1 transitions and for such transitions isospin symmetry does not 
predict equality of the reduced widths of analog transitions.
 
Isospin breaking Coulomb interaction plays a major role in "violating" these 
rules \cite{War}. For example, the energy of analog 
transitions can be affected by a Coulomb (iso-spin breaking) correction which is significant for 
quasi bound states (the Thomas-Ehrman shift \cite{Thomas}). Indeed it was shown that 
 such a Coulomb shift for the $1/2^+ \ \rightarrow \ 1/2^-$ analog transition in \yn13 
 amounts to a large fraction (25\%) of the transition energy. Barker \cite{Bar} evaluated the 
 Coulomb correction for the quasi bound state in \yn13, to yield a B(E1) which is more than 
 a factor of 3 different than its analog transition in $^{13}C$ (involving bound states). 
The $\Delta$T = 0 B(E1) in the $^{13}C$ - \yn13 iso-doublet are measured to 
be large fraction of the single particle Weisskopf Unit estimate, yet even for these 
strong analog transitions we observe large variations of the reduced widths.

In the case of weak transitions (as compared to single particle Weisskopf Unit estimate), the 
situation becomes even more confused. As pointed out by John Millener \cite{Millener}, 
for all multipolarities (except the purely isovector E1), the matrix
elements contain both isoscalar and isovector contributions so that
the reduced transition probabilities are of the form $(m_0 \ +\  m_1)^2$ and
$(m_0 \  -\  m_1)^2$ for the mirror transitions. For strong E2 transitions, $m_0$ dominates.
 For strong M1 transitions, $m_1$ dominates.
 For weaker transitions $m_0$ and $m_1$ are quite possibly  
comparable in magnitude, in which case the reduced transition
probabilities could be very different for the mirror transitions.
In addition cancellation are common for weak transitions which 
makes the theoretical situation very confusing, most 
notably for weak E1 transition. Estimates of the reduced widths of weak transitions may 
differ by several orders of magnitudes, since theory is not able to fine tune 
cancellation or the prediction of weak (E1) transitions.

High lying low spin states (e.g. high lying $0^+, 2^+$ states) that are most relevant for stellar 
burning, are well known to have complicated nuclear structure with admixture from several  
configurations. As such weak transitions originating from high lying low spin states are 
particularly hard to predict.

At last we note that bound states (e.g. the $0^+ _2$ in \ne22) and quasi bound states (e.g. 
the $0^+$ in \mg22) are expected \cite{War,Bar} to have wave function with different 
radial dependence which is expected to result considerable alteration of the width of 
mainly electric (but not magnetic) dipole transitions \cite{War,Bar}.

Based on these general comments one may not assume that the reduced widths of the analog 
transitions, and most certainly not the total radiative widths of analog states in \ne22 and \mg22 are equal, in sharp contrast to the assumption made by B.S. Davids {\em et al.} \cite{Dav1}.

As an explicit example we now consider the assumption \cite{Dav1} that the radiative width of the 
$0^+_2$ state at 5.962 MeV and the tentative $1^-$ states at 6.046 MeV in \mg22 are equal to the 
radiative width of the analog $0^+ _2$ state at 6.235 MeV and the analog $1^-$ state at 
6.689 MeV in \ne22. B.S. Davids {\em et al.} \cite{Dav1} use the measured radiative widths of 
these states  in \ne22 \cite{Endt,Ne22} to deduce the radiative widths in \mg22. 

We first note that no measured E1 transitions are 
involved in the decay of the $0^+_2$ in \ne22, hence even in the limit where iso-spin is an 
exact symmetry we do not expect the analog transitions in \mg22 to be equal as these are 
manifestly $\Delta$T = 0 transitions. Using the same data we extract in \ne22  \cite{Endt,Ne22} 
B(M1$\uparrow: \ 1+ \rightarrow 0^+ _2$) = 0.04 Wu. 
We conservatively estimate that the non-observed transition 
B(E2$\uparrow: \ 2^+ \rightarrow  0^+ _2$) could be as small as 0.001 Wu. 
This small B(E2) observed in \ne22 may arise from cancellations or a structure 
that is different than the ground state band in \ne22. In any case the analog E2 transition  
in \mg22 can not be assumed to have the exact same strength and be as small. 

If on the other hand one makes  the reasonable assumption that the analog transition 
B(E2$\uparrow: \ 2^+ \rightarrow  0^+ _2$) in \mg22 is of average 
strength of 1 Wu with an upper limit of 10 W.u. (as one may deduce from measured 
B(E2)s for the $0^+ _2$ in $^{18}O$, $^{18}Ne$, $^{20}Ne$ \cite{TUNL}), 
it will be the dominant electromagnetic decay mode with the largest electromagnetic 
branching ratio, due to its large transition energy. Such a transition will determine the total 
radiative width of the $0^+_2$ state in \mg22 to be very different than its analog $0^+ _2$ in \ne22.  
In such a case the estimated B(E2) will differ by a factor 
of 1000 with an upper limit of 10,000. 

Preliminary estimate of the 
B(E2$\uparrow: \ 2^+ \rightarrow  0^+ _2$) in \mg22 \cite{TRIUMF} with the hint that 
this transition is indeed the dominant decay of the $0^+_2$ in \mg22 \cite{TRIUMF}, yields  
approximately 0.3 W.u., and in this case B.S. Davids is in fact 
found to be wrong by a factor of approximately 300.

B.S. Davids {\em et al.} assume the analog of the 6.046 MeV 
state is the  $1^-$ at 6.689 MeV in \ne22, 
but it is quite possible that its analog is the $3^-$ state at 5.911 MeV in \ne22. In either case 
we observe for these states in \ne22 very weak B(E1) transitions 
ranging between 0.008 and 0.0002 W.u. For such weak E1s isospin symmetry can 
not guarantee the equality of the reduced widths of the analog transition in \mg22.

The cancellation discussed in this paper is reminiscent of the text book 
case of cancellation that leads to the anomalous 
long beta-decay lifetime of $^{14}C$ with log ft = 9.04 \cite{14C}. This small beta-decay matrix 
element is a million times retarded as compared to similar Gamow-Teller transitions in light nuclei 
with an average log ft = 3.5 \cite{GT}. The beta-decay matrix 
element in $^{14}C$ is a factor of 100 smaller than its analog in $^{14}O$ (log ft = 7.2), which 
emphasizes the fine tuning of matrix elements of analog states that involve cancellation.

Recent measurement of radiative widths in \mg22 \cite{TRIUMF} vividly illustrate how 
non useful are these very crude and mostly wrong guesses of radiative widths based 
on misapplication of isospin symmetry \cite{Dav1,Dav2,Dav3}. The measured central value of 
$\omega \gamma$ for the 5.962 MeV state differs by a factor of 4.3 and for the 6.046 MeV 
state it differs by 16.4. A comparison of Fig. 4 of B.S. Davids {\em et al.} 
\cite{Dav1} and Fig. 15 of Ref. \cite{TRIUMF} demonstrates that the burning rates at higher 
temperatures are very different than proposed or would be calculated based on the crude 
guesses of B.S. Davids {\em et al.}. Specifically the contribution of the 6.046 MeV state was 
considered by B.S. Davids {\em et al.} to be negligible and it was not 
included in the plot of contributions from "most important states" \cite{Dav1}. In fact this 
state contributes considerably more than the 5.962 MeV state shown in Fig. 4 \cite{Dav1}, 
and at high temperature its contribution competes with that of the 5.714 MeV that 
was considered by B.S. Davids {\em et al.} to be the only significant state. 
We also note that B.S. Davids restricted their calculation 
of nuclear burning to lower temperatures. Figure 15 of Ref. \cite{TRIUMF} is 
the correct one as it includes measured values of $\omega \gamma$ for states in 
\mg22, including the contribution of the 6.046 MeV state. This figure   
also includes the entire temperature range of interest for novae and x-ray bursts, unlike 
Fig. 4 of B.S. Davids {\em et al.} \cite{Dav1}.

The measurement of resonance strengths in \mg22 underlines the "danger" of (ab)using 
isospin symmetry to extract electromagnetic properties of analog transitions. Such 
practices are all too common in the field of Nuclear Astrophysics. The case 
discussed in \mg22 leads to conclusions that are off by large factors, large enough to make 
a difference even in Astrophysics.
 
 The author acknowledges very important and detailed comments from 
 John Millener who refreshed my memory concerning the rules included in Ref. 
 \cite{War} and the implication of the new measurement in $^{22}Mg$ \cite{TRIUMF}.

\end{document}